# Creating One-dimensional Nanoscale Periodic Ripples in a Continuous Mosaic Graphene Monolayer

Ke-Ke Bai[1,§], Yu Zhou[2,§], Hong Zheng[1], Lan Meng[2], Hailin Peng[2,*], Zhongfan Liu[2,*], Jia-Cai Nie[1], and Lin He[1,*]

[1] Department of Physics, Beijing Normal University, Beijing, 100875, People's Republic of China
[2] Center for Nanochemistry (CNC), College of Chemistry and Molecular Engineering, Peking University, Beijing, 100871, People's Republic of China

**In previous studies, it proved difficult to realize periodic graphene ripples with wavelengths of few nanometers. Here we show that one-dimensional (1D) periodic graphene ripples with wavelengths from 2 nm to tens of nanometers can be implemented in the intrinsic areas of a continuous mosaic (locally N-doped) graphene monolayer by simultaneously using both the thermal strain engineering and the anisotropic surface stress of Cu substrate. Our result indicates that the constraint imposed at the boundaries between the intrinsic and the N-doped regions play a vital role in creating these 1D ripples. We also demonstrate that the observed rippling modes are beyond the descriptions of continuum mechanics due to the decoupling of graphene's bending and tensional deformations. Scanning tunneling spectroscopy measurements indicate that the nanorippling generates a periodic electronic superlattice and opens a zero-energy gap of about 130 meV in graphene. This result may pave a facile way for tailoring the structures and electronic properties of graphene.**

∗ Correspondence to Lin He(helin@bnu.edu.cn) for general aspects of the paper, and to Hailin Peng(hlpeng@pku.edu.cn) and Zhongfan Liu(zfliu@pku.edu.cn) for sample fabrication.

§ These authors contribute equally to this work.

Graphene is a ultrasoft thin film and almost has no resistance against out-of-plane deformations [1,2]. Therefore, this unique one-atom-thick membrane is always wrinkled to certain degree [3,4]. Recently, much effort has been spent on accurate control of graphene ripples [2,5-7] motivated by the fact that ripple-induced lattice deformations can generate effective electric and magnetic fields on the propagating Dirac fermions, which strongly modify the electronic spectra and properties of graphene [5,6,8-14]. Therefore, strain engineering, which usually induces lattice distortions in graphene, becomes a powerful method for tuning the electronic properties of this ultimate thin film [2,5,6,9-13,15,16].

Importantly, graphene, being a ultrasoft thin membrane, offers unique advantages in this respect. It was demonstrated that one-dimensional (1D) periodic wrinkles with wavelengths $\lambda$ of several hundreds nanometers can be realized by thermal strain engineering and the classical membrane nature of graphene persists down to about 100 nm of the wrinkling wavelength [5,7]. Recently, Tapasztó, *et al.* observed periodic graphene ripples, with a wavelength of only 0.7 nm and modulation of only 0.1 nm, over the trenches with widths of about 5 nm in Cu(111) surface [2]. Their result indicates the breakdown of plate phenomenology (i.e., the continuum mechanics) for this 0.7 nm ripples of graphene. A natural question then arises: what's the behavior of the 1D periodic graphene ripples with 0.7 nm $< \lambda <$ 100 nm? Unfortunately, rare experiment realizes the periodic graphene ripples with wavelength in such a region. According to the continuum mechanics, the rippling wavelength scales with the square root of the membrane size. It implies that one should apply a strain on a nanometer-sized suspended graphene to implement nanoscale periodic graphene ripples [2]. This is, to some extent, out of the grasp of today's technology and impedes investigation of graphene ripples down to this length scale. Therefore, a validity-check of the continuum mechanics for the nanoscale graphene ripples is still missing up to now.

In this Letter, we show that such graphene ripples can be realized during the chemical vapour deposition growth of modulation-doped mosaic graphene, a continuous graphene monolayer with regionally varied doping profile [17], on Cu foils. The periodic graphene ripples with wavelengths from 2 nm to several tens nanometers are observed using scanning tunneling microscopy (STM). This enables unprecedented experimental access to the nanoscale periodic graphene ripples and provides an idea platform to exploit the effect of nano-rippling on the electronic spectra of graphene. Our experiment demonstrates that the ripples generate a periodic electronic superlattice and open a sizable band gap in graphene.

In our experiment, there are two steps to synthesize the mosaic graphene [17]. First, discrete intrinsic graphene grains were grown on the Cu substrate with methane vapour (step one). After a short purging period, these graphene grains serve as matrices for the laterally grafted growth of the N-doped graphene with acetonitrile vapour (step two) (see supporting materials [18] for details). Coalescence of the graphene grains yields a continuous mosaic graphene monolayer. The spatially well-defined intrinsic (i) and

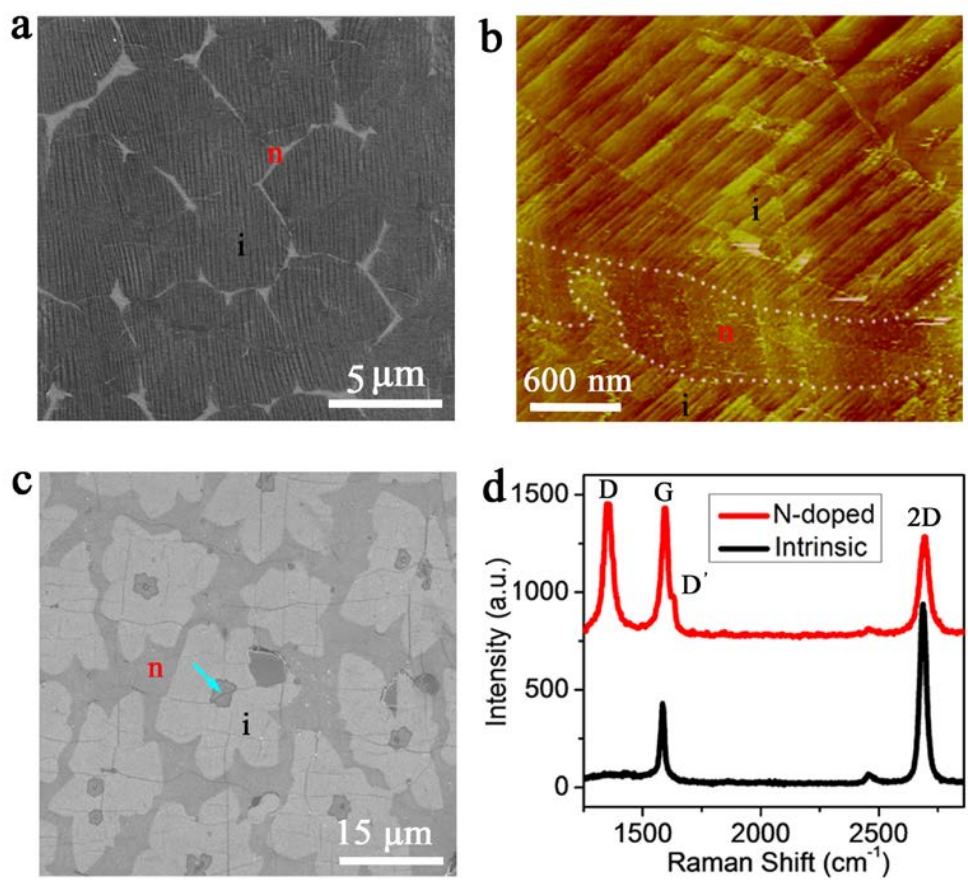

**Figure 1 (color online).** Morphology of modulation-doped graphene. **a**, A SEM image of a mosaic graphene on Cu foils. Intrinsic (i) and N-doped (n) portions can be identified by the contrast. **b**, A scanning tunnelling microscopy image of a mosaic graphene measured at atmosphene. The white dotted curves indicate the boundaries between the intrinsic and N-doped portions of the mosaic graphene. **c**, A SEM image of mosaic graphene transferred onto a 300 nm SiO$_2$/Si substrate. The light blue arrow indicates the bilayer or multilayer graphene. **d**, Typical Raman spectra of intrinsic and N-doped portions of mosaic graphene collected with a 514 nm incident laser. The black curve exhibited sharp G and 2D Raman bands, with the ratio of I$_{2D}$/I$_G$ > 2 was identified as the intrinsic graphene. The absence of D Raman band indicates the high quality of graphene with few of defects. In contrast, the red curve exhibited strong D and D' bands with broaden and shift of both G and 2D bands was recognized as the N-doped portion, indicating the lower quality of the direct growth of N-doped graphene.

N-doped (n) portions of the mosaic graphene show alternating contrast in the scanning electron microscopy (SEM) images and also show different Raman spectra, as shown in Fig. 1. The SEM investigation of the as-grown sample shows the presence of 1D periodic wrinkles with wavelength of hundreds of nanometers (Fig. 1a). During the cooling process, mismatch of thermal expansion coefficients between graphene and the copper foil leads to the formation of wrinkles. The origin of the 1D compression is mainly attributed to the anisotropic surface stress of the Cu substrate [7]. Besides the large graphene wrinkles, nanorippling modes with wavelengths ranging from only several nanometers to tens of nanometers are also observed in the intrinsic graphene around the boudaries between the intrinsic and N-doped portions of the mosaic graphene, as shown in Fig. 1b (and Fig. S1 of supporting materials [18]). The length of these ripples is typically in the range of 400-1000 nm.

Previously, it was believed that nanoscale periodic graphene ripples can only be implemented in nanometer-sized graphene [2,5,7,19], and no one actually expected they to exist in a continuous graphene monolayer. Obviously, the observed 1D nanoscale periodic graphene ripples in the mosaic graphene monolayer violate the predictions of the continuum model and are contrary to current expectations. Here we should point out that the rippling pattern in the mosaic graphene is distinct from that of only intrinsic graphene flakes grown on Cu foils (or on liquid Cu) [7,20]. To further explore the effect of the N-doped regions on the emergence of these nanoscale ripples, two intrinsic graphene samples grown on Cu foils were synthesized for comparison. One is the discrete intrinsic graphene grains obtained in the process of the "step one". The other is a continuous intrinsic graphene monolayer synthesized by a similar process as that of the mosaic graphene monolayer. The only difference is that the methane vapour rather than the acetonitrile vapour is used in the step two to obtain the continuous intrinsic graphene monolayer (see supporting materials [18] for details). For the two intrinsic graphene samples, only large periodic wrinkles with wavelength of hundreds of nanometers can be observed (see Fig. S2 of supporting materials [18]). Additionally, the large periodic wrinkles appear in every part of the continuous intrinsic graphene. This differs much from that of the modulation-doped mosaic graphene, in which the rippling is not present in the N-doped regions.

The above controlled experiments indicate that the nanorippling mode in the mosaic graphene is closely related to the absence of rippling in the N-doped regions. The constraint imposed at the boudaries between the intrinsic and N-doped portions of the mosaic graphene may be the main reason for the emergence of the nanoscale ripples. A clear evidence is that these nanoscale ripples only appear beside the boundaries. According to the result of the continuous intrinsic graphene monolayer, it is reasonable to conclude that the N dopants play a key role for the absence of rippling in the N-doped regions. In the N-doped portion of the mosaic graphene, the STM images show the graphene lattice with superimposed bright objects, which are attributed to the N dopants (see Fig. S3 of

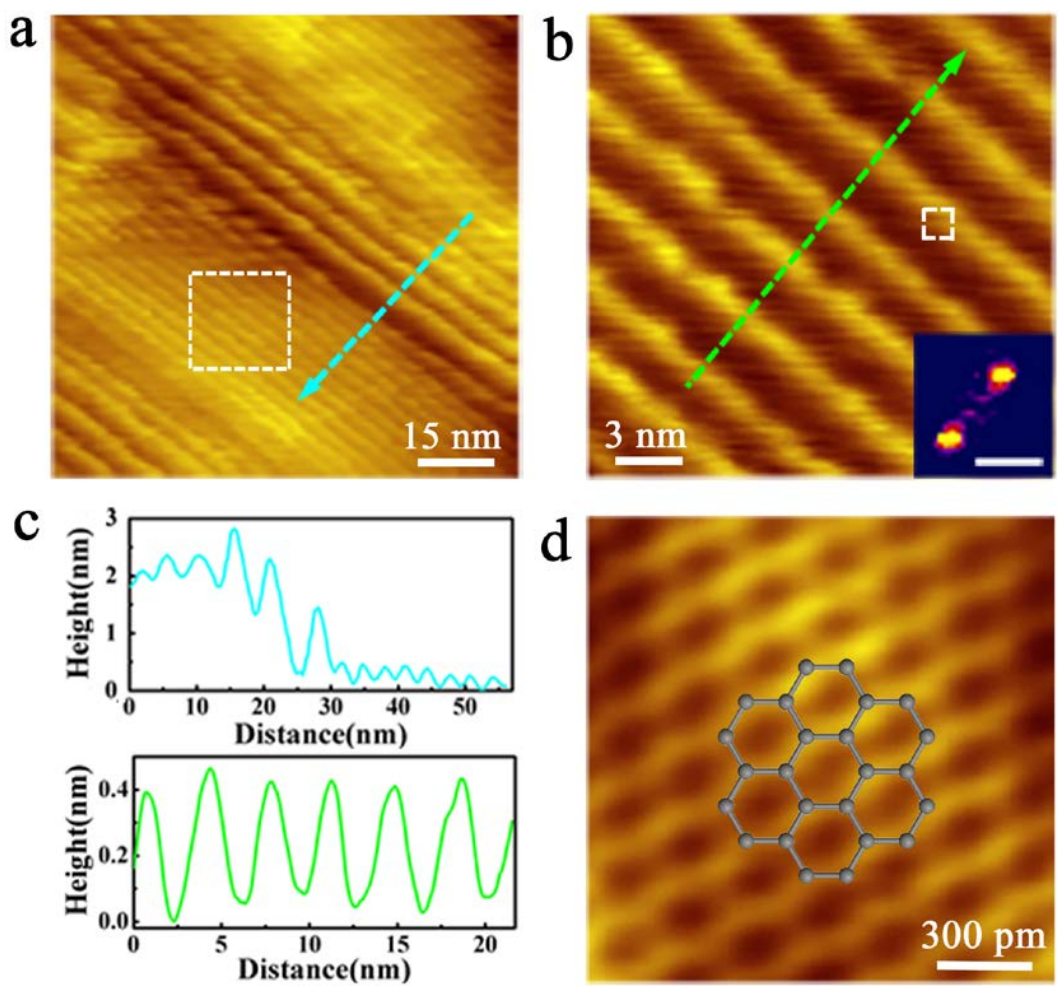

**Figure 2 (color online).** Atomic-resolution STM images of graphene nanoripples. **a**, A STM topography of a graphene sheet with nanoscale graphene ripples crossing over a copper step ($V_{sample}$ = 0.62 V and $I$ = 19.2 pA). **b**, Zoom-in topography of the white frame in panel (**a**) ($V_{sample}$ = 0.62 V and $I$ = 28.6 pA). The inset is the Fast Fourier transform showing the reciprocal-lattice of the periodic ripples. The scale bar is 3 Gm$^{-1}$. **c**, the upper panel: the height profile along the dashed line in (**a**); the lower panel: the height profile along the dashed line in (**b**). **d**, Atomic-resolution STM image of the white frame in panel (**b**) ($V_{sample}$ = 0.79 V and $I$ = 35.8 pA ) exhibiting the honeycomb lattice of graphene. The atomic structure of graphene is overlaid onto the STM image.

supporting materials [18] for a typical STM image) [21]. These N dopants can be treated as point defects, which generate large strain and lattice distortion around themselves [21-25]. Our result suggests that the graphene with high density of point defects is not inclined to form ripples for strain relaxation when it is subjected to a compression induced by the substrate. It may through lattice distortion around the N dopants to relax the in-plane compressive strains. More theoretical studies should be carried out to quantitative understand the effect of the N dopants and the role of the boundary on the appearance of the nanoscale rippling.

The atomic structure of the nanoscale graphene ripples was studied using high-resolution STM. A typical STM topograph of a corrugated area is shown in Fig. 2a. To validate the structural origin of the periodic ripples rather than the electron interference, we used different bias voltage to acquire the STM images and the observed pattern was almost unchanged. Continuity of flat graphene over steps of copper surface has been demonstrated previously by STM studies [26]. Our result indicates that such a continuity is still valid even when the graphene monolayer is periodic corrugated (see Fig. 2a and the upper panel of Fig. 2c). The aperiodic structure of the graphene ripple around the step also excludes the electron interference as its origin. The STM image in Fig. 2b revealed that the local periodicity of the nanoscale ripples above a Cu terrace could be quite uniform. The height field of the graphene ripples (as shown in the lower panel of Fig. 2c) can be approximated well by a sinusoidal function $H(x) = A\sin(2\pi x/\lambda)$ with the amplitude $A \sim 0.2$ nm and the

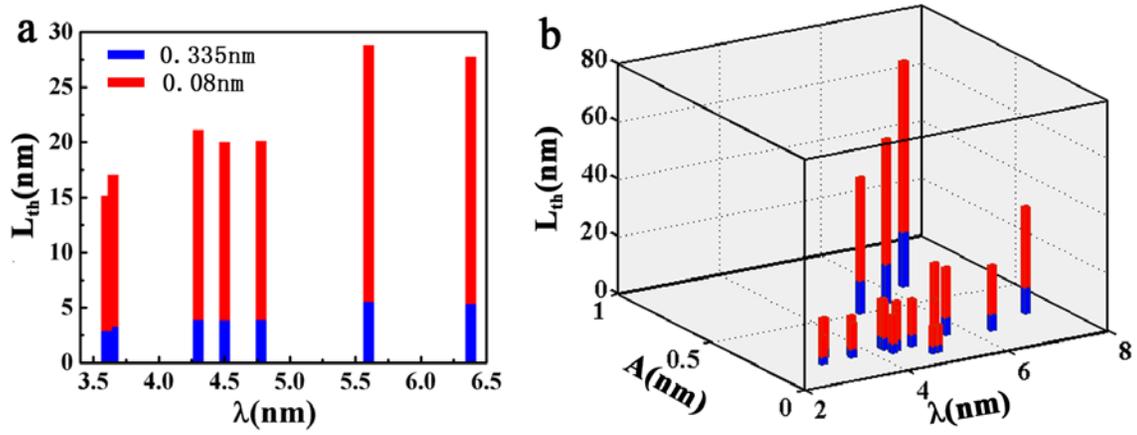

**Figure 3 (color online) a**, The theoretical length ($L_{th}$) as a function of wavelength for $A = 0.2 \pm 0.02$ nm. The $L_{th}$ is calculated according to Eq. (1) with $t = 0.08$ nm (red bars) and $t = 0.335$ nm (blue bars) respectively. **b**, The theoretical lengths, experimental wavelengths, and experimental amplitudes of all the graphene nanoripples observed in our experiments. The wavelength $\lambda$ and the amplitude $A$ of the nanoscale ripples are determined from STM images in the experiments.
9

period $\lambda \sim 3.2$ nm. Atomic resolution image taken in the nanorippling areas (Fig. 2d) shows a perfect honeycomb lattice of graphene, which suggests that the small local curvature of the ripples doesn't destroy the sixfold symmetry of the graphene lattice. The lacking of N dopants in the nanorippling areas (see Fig. S4 of supporting materials [18] for more experimental data) further confirms the result that the nanoscale ripples only appear in the intrinsic portion of the mosaic graphene.

The ability to ripple graphene down to nanometer wavelengths, as reported here, allows us to explore the rippling of this one atom thick membrane beyond the description of continuum mechanics. According to the continuum mechanics, the length $L$ of the ripples is determined by an equation with only experimentally accessible parameters [2,5,7]:

$$L = \frac{A\lambda}{\sqrt{8v/(3(1-v^2))t}} \quad . \tag{1}$$

Here, $t$ is the thickness of the membrane and $v$ is the Poisson ratio [2,5]. For single-layer graphene, there is the ambiguity of defining the thickness of this single-atom-thick membrane of carbon [2,27]. Usually, $t = 0.335$ nm, the experimentally measured interlayer spacing in graphite, and $t = 0.08$ nm, the effective thickness derived from atomistical theory [27], are widespread used to describe the mechanics of graphene monolayer.

We can immediately estimate the theoretical length of the graphene ripples as $L_{th} \sim 2.9$ nm according to Eq. (1) using $A = 0.2$ nm, $\lambda = 3.2$ nm (the amplitude and wavelength of the ripples are determined from Fig. 2b), $v = 0.16$, and $t = 0.335$ nm. Obviously, this value is

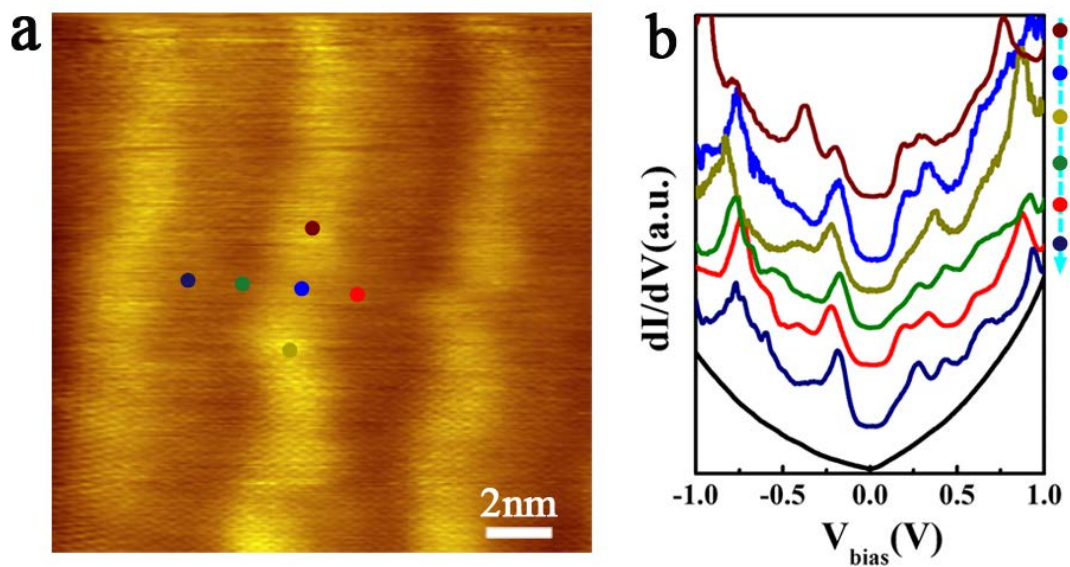

**Figure 4 (color online)** The STM images and STS of graphene nanoripples. **a**, A STM topography of a periodic rippled graphene on a Cu foil ($V_{sample}$ = 0.67 V and $I$ = 69.4 pA). **b**, Tunnelling spectra, i.e., dI/dV-V curves, recorded at different positions, as marked by the dots with different colors in panel (**b**), of the graphene ripples. The spectra were vertically offset for clarity. A spectrum recorded on flat graphene area (black curve) grown on the Cu foil is also shown for comparison.

more than two orders of magnitude lower than our experimental result, as shown in Fig. 2. Varying the thickness of graphene from $t = 0.335$ nm to $t = 0.08$ nm, the theoretical length of the ripples is changed to $L_{th} \sim 12.1$ nm, which still fails by more than an order of magnitude. It implies that the phenomenological continuum model is not valid to describe the nanorippling observed in our experiments. This result is also verified for other nanoscale graphene ripples with wavelengths of several nanometers, as shown in Fig. 3. The theoretical lengths of the ripples calculated with both $t = 0.335$ nm and $t = 0.08$ nm are much smaller than our experimental result, $400$ nm $< L_{exp} < 1000$ nm. Such a breakdown of the continuum mechanics in graphene monolayer is reasonable since that the resistance to bending in graphene only involves the $\pi$-orbital misalignment between adjacent pairs of C atoms, which implies almost zero resistance against out-of-plane deformations [1,2]. Therefore, graphene can ripple at the nanometer and even subnanometer scale. In a classical plate depicted by the continuum mechanics, the bending always induces the in-plane stretching and compression on the opposite sides of a neutral curved surface, which generates large resistance against the out-of-plane deformations.

The observed nanoscale ripples in graphene are more than just structural curiosity, since they give rise to electronic superlattices [2,15,28], which are predicted to affect the electronic properties of graphene dramatically [29-32]. We performed scanning tunneling spectroscopy (STS) measurements to explore the effect of the periodic nanoscale rippling on the local electronic properties of the graphene. Figure 4 shows several typical STS spectra recorded at different positions of a 1D periodic graphene ripples with the amplitude $A \sim 0.2$ nm and the period $\lambda \sim 4.7$ nm. A STS spectrum obtained on flat area of graphene on the copper foil is also shown for comparison. The tunneling spectrum gives direct access to the local density of electronic states of the surface at the position of the STM tip. At low bias voltages, a band gap ~ 130 meV with almost zero conductance is clearly observed for the graphene nanoripples. Very recently, it was demonstrated by transport measurements that the electronic superlattices can induce a band gap in graphene monolayer [6,33,34]. A band gap within 0.14-0.19 eV is observed in a periodically modulated graphene and the origin of the gap is attributed to both the periodic mechanical bending and the surface coordination chemistry [6]. Many other factors, such as the effect of substrate and the enhanced electron-electron interactions, could contribute to the observed gap [6,32-34]. In strained graphene, the coupling between the strain-induced pseudomagnetic field and a scalar electric potential is also predicted to generate a significant energy gap [35]. Besides the band gap, two dips of the local density of states at $V_{bias} \sim \pm 0.9$ eV, which are attributed to the minigaps opened by Bragg scattering at principal superlattice harmonics [32], are observed. The minigaps generated by the ripple-induced electronic superlattice should appear at $\pm \hbar v_F|\mathbf{G}|/2 = \pm 0.9$ eV (here $v_F$ is the Fermi velocity of graphene and $\mathbf{G}$ is the reciprocal electronic superlattice vector), which agrees quite well with our experimental result. The result in Fig. 4 provides direct evidence that nanoscale rippling in graphene could generate 1D electronic superlattices and further induce a sizable band gap, which is urgent to enable graphene's technological adoption.

In summary, we demonstrate that 1D nanoscale periodic ripples can be implemented in a continuous mosaic graphene. The observed ripples violate the predictions of the continuum model and are contrary to current expectations. Our experimental result indicates that the ripple generates a periodic electronic superlattice and induces a sizable band gap in graphene. Therefore, the 1D nanoscale periodic graphene ripples are important not only in fundamental physics, but also in advanced nanoelectronics.

**Acknowledgements**

This work was supported by the National Basic Research Program of China (Grants Nos. 2014CB920903, 2013CBA01603, 2013CB921701), the National Natural Science Foundation of China (Grant Nos. 11374035, 11004010, 51172029, 91121012), the program for New Century Excellent Talents in University of the Ministry of Education of China (Grant No. NCET-13-0054), and Beijing Higher Education Young Elite Teacher Project (Grant No. YETP0238).